# Airtight metallic sealing at room temperature under small mechanical pressure

Stephen P. Stagon and Hanchen Huang*

Department of Mechanical Engineering, University of Connecticut, Storrs, CT 06269, USA

Metallic seals can be resistant to air leakage, resistant to degradation under heat, and capable of carrying mechanical loads. Various technologies – such as organic solar cells and organic light emitting diodes – need, at least benefit from, such metallic seals [1-7]. However, these technologies involve polymeric materials and do not tolerate the either high-temperature or high-pressure processes of conventional metallic sealing. Recent progress in nanorod growth [8] opens the door to metallic sealing to these technologies. Here, we report a process of metallic sealing using small well-separated Ag nanorods; the process is at room temperature under a small mechanical pressure of 9.0 MPa, and also in ambient. The metallic seals have an air leak rate of $1.1 \times 10^{-3}$ cm$^3$atm/m$^2$/day, and mechanical shear strength higher than 8.9 MPa. This leak rate meets the requirement of organic solar cells and organic light emitting diodes [1].

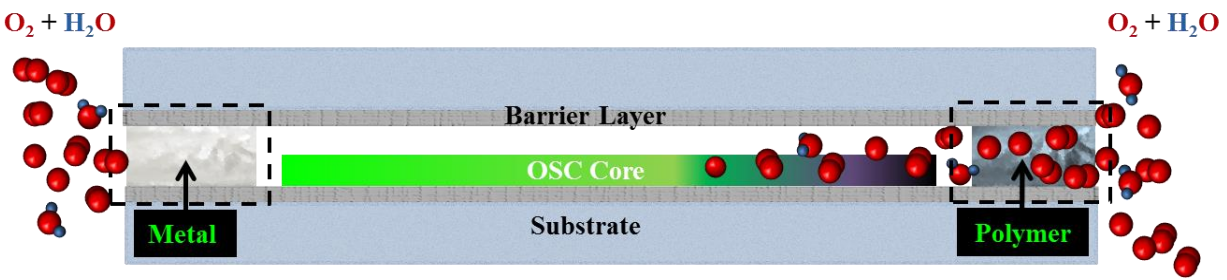

**Figure 1: Need of metallic sealing.** Schematic of air, particularly oxygen (O$_2$) and water vapor (H$_2$O), leak in OSCs.

To begin, we first use organic solar cells (OSCs) as an example to demonstrate the challenges of sealing and identify the need of metallic sealing at room temperature, under small mechanical pressure, and preferably in ambient environment. As shown in Figure 1, an organic semiconductor core of an OSC is encapsulated to avoid exposure to air, particularly oxygen ($O_2$) and water vapor ($H_2O$). The exposure leads to degradation and short lifetime [3,4]. For flexibility, the encapsulating substrates are usually polymers and have barrier layers on them to minimize the exposure. At the same time, the seals that connect the substrate and the barrier layer must be airtight to minimize the exposure. At present, the seals are polymeric, as shown on the right of Figure 1, and they do not provide sufficient resistance to air leakage, even when new. During operation, the polymer seals degrade and their air leak rate increases even more. As a result, the OSCs have short lifetime, which is one of the main reasons that OSCs are not yet economically competitive [4,6]. Similarly, organic light emitting diodes (OLEDs) have apparent technological advantage but their wide spread implementation has not been possible due to the exposure and short lifetime [6]. The metallic seal, shown on the left of Figure 1, can be sufficiently airtight. However, metallic seals are absent in OSCs or OLEDs despite their apparent advantage in air leak resistance. This absence is the result of a major technical challenge – the organic semiconductor core and the polymeric substrates are incompatible with sealing processes at high temperature or high pressure.

Next, we examine the existing processes of metallic sealing, with particular focus on processing temperature and pressure. Since the sealing temperature must be low to avoid damage to the organic semiconductor core and the polymer substrates, our examination focuses on only those processes at relatively lower temperatures. The existing processes fall into three categories. First, the soldering process that has been commonly used in the electronics industry functions much above 60°C with eutectic alloys containing toxic Pb or expensive In [9-11]. Even without any consideration of toxicity and cost, such alloys that allow soldering around 60°C have low mechanical strength during normal OSC operation which easily approaches their melting temperature. Further, the soldering process involves corrosive fluxes and the use of vacuum. Second, cold welding of flat surfaces functions at room temperature, but it requires an extremely high compressive load, on the order of about 1 GPa [12]. However, polymeric substrates may disfigure even at a compressive load that is one order of magnitude lower, about 100 MPa [13]. Third, sealing with metallic nanoparticles or nanorods takes advantage of fast surface diffusion and occurs at relatively low temperatures [14]. Using Ag nanoparticles, the sealing process is possible at 160°C through coarsening under a low pressure of about 10 MPa [15]. The processing temperature, which is too high, cannot be lowered because of the organic capping layer on each nanoparticle that originates from solution synthesis. While it is possible to remove the capping layer with solvents, such solvents are not compatible with the organic components of the OSCs and OLEDs [16]. With not-well-separated Cu nanorods, the sealing process requires a high temperature of 300°C under low pressure in a vacuum or inert environment [17]. Even without the consideration of the processing cost associated with a non-ambient environment, none of the three categories of seals satisfy the requirements of low temperature and low pressure for OSC and OLED technologies.

    Here, we propose a metallic sealing process at room temperature under a small mechanical pressure, below 10 MPa, and also in ambient environment. This proposal builds on the growth of small well-separated metallic nanorods, an ability that has recently become reality

[8,18]. These nanorods do not have capping layers, and do not require a high temperature of 160°C to coarsen. Further, we choose Ag for metallic sealing to minimize potential oxidation with only moderate cost; in comparison, the volatile and increasing cost of In in eutectic alloys for soldering is greater than that of Ag [9-11,19]. To ensure good adhesion, we add a metallic film between the Ag nanorods and sealing substrate.

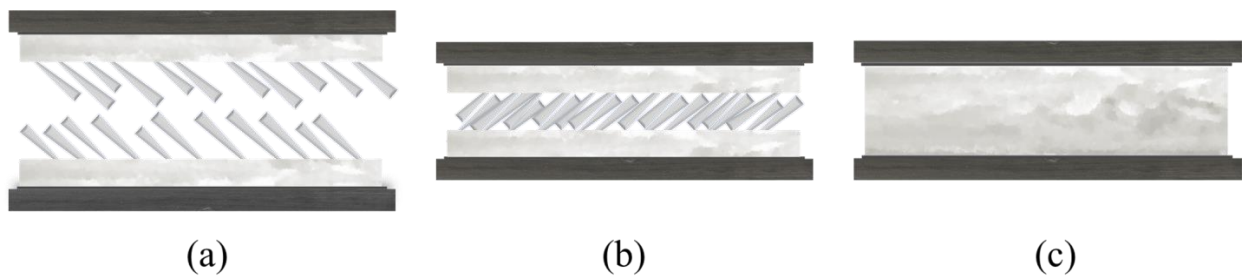

**Figure 2: Schematic of metallic nanorod sealing.** Schematics of metallic nanorods (gray) on metallic film (gray) and substrate (dark), (a) before, (b) during, and (c) after the sealing process.

Figure 2 schematically illustrates our proposal. Before the sealing process, two substrates carrying the small well-separated Ag nanorods are brought to face each other; Figure 2a. Under a small mechanical pressure, the nanorods from two sides crosslink with each other; Figure 2b. The feasibility of crosslinking is critical in this proposal, and it benefits from the recent realization of small well-separated nanorods [8]. Due to the fast diffusion on the surfaces of nanorods, the crosslinked nanorods condense into a film; Figure 2c.

The fast surface diffusion is necessary for the proposed sealing process in Figure 2, and it is next verified. Figure 3 shows the morphological change of Ag nanorods during annealing in ambient at constant temperatures; see Supplemental Materials Sections S2 and S3 for details of fabrication, annealing, and characterization of the nanorods. Figure 3a shows the Ag nanorods that are kept at room temperature for about one hour from their synthesis. The bridging between nanorods indicates, non-conclusively, potentially fast surface diffusion even at room temperature, which is nominally 25°C. At 50.0°C ± 4°C, only slightly above room temperature, the substantial change of morphologies over merely five minutes shows that the surface diffusion is indeed fast; Figure 3b. At 75.0°C ± 6°C, the change of morphologies over five minutes is more dramatic, indicating even faster surface diffusion; Figure 3c. At 100.0°C ± 8°C, the nanorods completely coalesce into a continuous film in five minutes; Figure 3d. This set of annealing results confirms that surface diffusion near room temperature can be fast, and that surface diffusion slightly above room temperature is very fast.

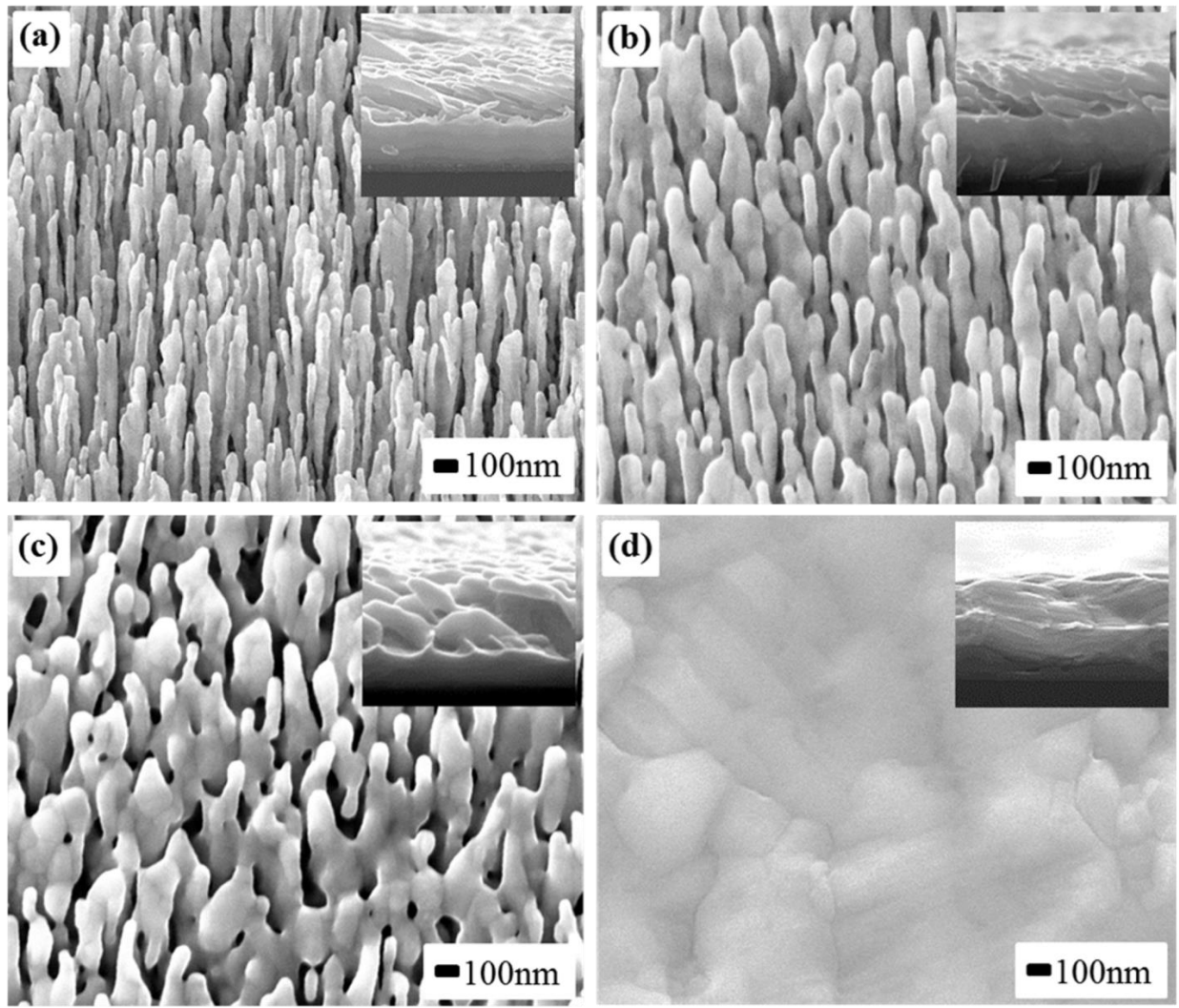

**Figure 3: Ag nanorod annealing.** SEM images of Ag nanorods (a) before annealing from a top view, with 2° imaging titled cross-section view as inset; and after annealing for five minutes at (b) 50°C, (c) 75°C, and (d) 100°C.

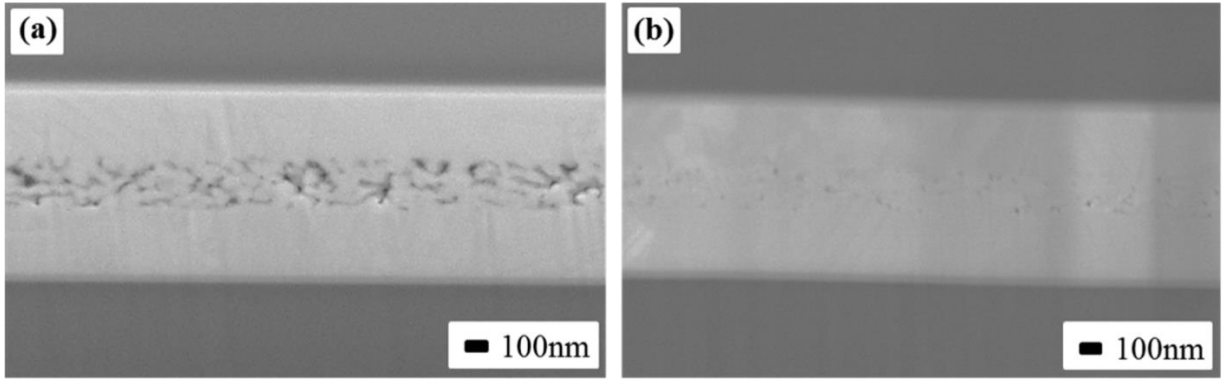

**Figure 4: Seal morphologies.** Cross-sectional SEM images of seals between two substrates under a small mechanical pressure of 9.0 MPa in ambient for five minutes (a) at room temperature, and (b) at 100°C.

Based on the annealing results, we first test the proposed metallic sealing at room temperature. As Figure 4a shows, the seal formed at room temperature under a small pressure of 9.0 MPa ± 1.3 MPa for five minutes consists of continuous solid regions; a longer compression time of 30 minutes leads to visually the same seal. Details of the sealing process are available in Supplemental Materials section S4. Although voids exist, the two substrates are well connected, leaving no apparent gap. The absence of gaps will provide the leak resistance, as our measurements of air leak rate will confirm later. In contrast, earlier attempts using Cu nanorods resulted in metallic seals with an apparent gap between the two substrates, even at much higher sealing temperatures of 200°C and 300°C [17]. To test the effects of faster surface diffusion, we have repeated the sealing process at 100°C $\pm$ 8°C for five minutes. As shown in Figure 4b, sealing at 100°C essentially eliminates voids beyond the size of a few nanometers.

Going beyond the morphologies of the seals, we now put the seal of Figure 4a to the test for air leak; see section S4 of the Supplemental Materials for details of the setup and measurements. According to direct measurement of pressure degradation as a function of time inside a sealed vacuum, we determine the air leak rate to be less than $6.7 \times 10^{-10}$ cm$^3$atm/s, taking into account a very conservative error bar. To appreciate how small this leak rate is, we compare it with (1) the leak rate of polymeric adhesive, and (2) the desired standards of the OSC and OLED industries. First, repeating the leak test with polymeric glue, we determine the leak rate to be at least 1000 times higher than that of the metallic seal. Second, when it comes to the industry standard, the requirements of leak resistance are $1 \times 10^{-3}$ cm$^3$atm/m$^2$/day for $O_2$ to $1 \times 10^{-4}$ cm$^3$atm/m$^2$/day for $H_2O$ vapor, for a reference configuration of 1 m x 1 m square solar panel [1]. For such a reference configuration, the air leak rate of our metallic seal is equivalently $1.5 \times 10^{-3}$ cm$^3$atm/m$^2$/day. Considering that 21% of typical air is $O_2$ and 3% is $H_2O$ vapor (in volume), the corresponding leak rate of $O_2$ is $3.2 \times 10^{-4}$ cm$^3$ atm/m$^2$/day and that of $H_2O$ vapor is $4.5 \times 10^{-5}$ cm$^3$ atm/m$^2$/day [20]. These are several times better than the industry requirements for both $O_2$ and $H_2O$ vapor [1]. We note that this better-than-required leak rate is achieved at room temperature under small mechanical pressure of 9.0 MPa, and also in ambient environment. Since the seal from room temperature processing suffices, here we will not pursue the test of seals from higher temperature processing.

As an additional step, we have examined the mechanical shear strength of the metallic seal. Using lap shear pull tests, we determine the lower limit of the shear strength of the seal in Figure 4a to be 8.9 MPa; see Supplemental S5 for details of measurements. Repeating the tests using seals formed under mechanical pressure of about 5 MPa, we find that the air leak rate does not change by more than 10% but mechanical delamination occurs between the seal and the polymeric substrates. That is, for both air leak resistance and mechanical strength, the mechanical compression of up to about 10 MPa is appropriate.

In summary, we report a metallic sealing process at room temperature under small mechanical pressure of 9.0 MPa, and also in ambient environment, for the first time. Through the easily accessible process, the resulting metallic seal has an air leak rate that is 1000 times better than that of polymeric glue, and several times better than that desired by the OSC and OLED industries [1]. Multiple technologies – such as OSCs and OLEDs – will benefit from this metallic sealing process.

**References:**

1. Krebs, F. C. & Norrman, K. Analysis of the failure mechanism for a stable organic photovoltaic during 10,000 h of testing. *Prog. Photovoltaics Res. Appl.* **15**, 697-712 (2007).

2. Forrest, S. R. The path to ubiquitous and low-cost organic electronic appliances on plastic. *Nature* **428**, 911-918 (2004).

3. Kawano, K. *et al*. Degradation of organic solar cells due to air exposure. *Solar Energy Mater. Solar Cells* **90**, 3520-3530 (2006).

4. Dennler, G. *et al*. A new encapsulation solution for flexible organic solar cells. *Thin Solid Films* **511**, 349-353 (2006).

5. Burrows, P. *et al*. Reliability and degradation of organic light emitting devices. *Appl. Phys. Lett.* **65**, 2922-2924 (1994).

6. Lewis, J. Material challenge for flexible organic devices. *Mater. Today* **9**, 38-45 (2006).

7. Reineke, S. *et al*. White organic light-emitting diodes with fluorescent tube efficiency. *Nature* **459**, 234-238 (2009).

8. Niu, X. B., Stagon, S. P., Huang, H. C., Baldwin, J. K. & Misra, A. Smallest metallic nanorods using physical vapor deposition. *Phys. Rev. Lett.* **110**, 136102 (2013).

9. Ma, H. & Suhling, J. C. A review of mechanical properties of lead-free solders for electronic packaging. *J. Mater. Sci.* **44**, 1141-1158 (2009).

10. Rathmell, A. R., Bergin, S. M., Hua, Y., Li, Z. & Wiley, B. J. The growth mechanism of copper nanowires and their properties in flexible, transparent conducting films. *Adv. Mater* **22**, 3558-3563 (2010).

11. Tolcin, A. Indium U.S. Geological Survey. *USGS Mineral Commodity Summaries* (2013).

12. Akande, W. O., Cao, Y., Yao, N. & Soboyejo, W. Adhesion and the cold welding of gold-silver thin films. *J. Appl. Phys.* **107**, 043519 (2010).

13. Abu-Isa, I. A., Jaynes, C. B. & O'Gara, J. F. High-impact-strength poly (ethylene terephthalate)(PET) from virgin and recycled resins. *J. Appl. Polym. Sci.* **59**, 1957-1971 (1996).

14. Lu, Y., Huang, J. Y., Wang, C., Sun, S. & Lou, J. Cold welding of ultrathin gold nanowires. *Nat. Nanotechnol.* **5**, 218-224 (2010).

15. Alarifi, H., Hu, A., Yavuz, M. & Zhou, Y. N. Silver nanoparticle paste for low-temperature bonding of copper. *J. Electron Mater.* **40**, 1394-1402 (2011).

16. Magdassi, S., Grouchko, M., Berezin, O. & Kamyshny, A. Triggering the sintering of silver nanoparticles at room temperature. *ACS Nano* **4**, 1943-1948 (2010).

17. Wang, P. *et al*. Low temperature wafer bonding by copper nanorod array. *Electrochem. Solid-State Lett.* **12**, H138-H141 (2009).

18. Huang, H. C. A framework of growing crystalline nanorods. *JOM* **64**, 1253 (2012).

19. Michael, G. Silver U.S. Geological Survey. *USGS Mineral Commodity Summaries* (2013).

20. Brimblecombe, P. in *Air composition and chemistry Ch. 1* (Cambridge University Press, New York, 1996).

**Acknowledgements:** The authors acknowledge the financial support of US DoE Office of Basic Energy Science (DE-FG02-09ER46562). They also thank Roger Ristau for assistance in FIB milling experiments, and Paul Elliot for assistance in PVD experiments.

**Contributions:** The two authors designed the project, analyzed the results, and wrote the manuscript together; and SPS performed the experiments.

**Additional information:** Supplementary information is available in the online version of the paper. Reprints and permissions information is available online at www.nature.com/reprints. Correspondence and requests for materials should be addressed to HCH.

**Competition financial interest:** The authors declare no competing financial interests.